\documentclass[preprint2,longabstract]{aastex}

\shorttitle{Pulsed Gamma-rays from the millisecond pulsar J0030+0451 with the \emph{Fermi} Large Area Telescope}
\shortauthors{Fermi-LAT Collaboration}
\begin{document}

\title{Pulsed Gamma-rays from the millisecond pulsar J0030+0451 with the \emph{Fermi} Large Area Telescope}

\author{
A.~A.~Abdo\altaffilmark{1,2}, 
M.~Ackermann\altaffilmark{3}, 
W.~B.~Atwood\altaffilmark{4}, 
M.~Axelsson\altaffilmark{5,6}, 
L.~Baldini\altaffilmark{7}, 
J.~Ballet\altaffilmark{8}, 
G.~Barbiellini\altaffilmark{9,10}, 
D.~Bastieri\altaffilmark{11,12}, 
M.~Battelino\altaffilmark{5,13}, 
B.~M.~Baughman\altaffilmark{14}
% ...
}
\altaffiltext{1}{National Research Council Research Associate}
\altaffiltext{2}{Space Science Division, Naval Research Laboratory, Washington, DC 20375}
\altaffiltext{3}{W. W. Hansen Experimental Physics Laboratory, Kavli Institute for Particle Astrophysics and Cosmology, Department of Physics and SLAC National Laboratory, Stanford University, Stanford, CA 94305}
\altaffiltext{4}{Santa Cruz Institute for Particle Physics, Department of Physics and Department of Astronomy and Astrophysics, University of California at Santa Cruz, Santa Cruz, CA 95064}
\altaffiltext{5}{The Oskar Klein Centre for Cosmo Particle Physics, AlbaNova, SE-106 91 Stockholm, Sweden}
\altaffiltext{6}{Department of Astronomy, Stockholm University, SE-106 91 Stockholm, Sweden}
\altaffiltext{7}{Istituto Nazionale di Fisica Nucleare, Sezione di Pisa, I-56127 Pisa, Italy}
\altaffiltext{8}{Laboratoire AIM, CEA-IRFU/CNRS/Universit\'e Paris Diderot, Service d'Astrophysique, CEA Saclay, 91191 Gif sur Yvette, France}
\altaffiltext{9}{Istituto Nazionale di Fisica Nucleare, Sezione di Trieste, I-34127 Trieste, Italy}
\altaffiltext{10}{Dipartimento di Fisica, Universit\`a di Trieste, I-34127 Trieste, Italy}
\altaffiltext{11}{Istituto Nazionale di Fisica Nucleare, Sezione di Padova, I-35131 Padova, Italy}
\altaffiltext{12}{Dipartimento di Fisica ``G. Galilei", Universit\`a di Padova, I-35131 Padova, Italy}
\altaffiltext{13}{Department of Physics, Royal Institute of Technology (KTH), AlbaNova, SE-106 91 Stockholm, Sweden}
\altaffiltext{14}{Department of Physics, Center for Cosmology and Astro-Particle Physics, The Ohio State University, Columbus, OH 43210}
% ...

\begin{abstract} 
We report the discovery of gamma-ray pulsations from the nearby isolated millisecond pulsar PSR J0030+0451 with the Large Area Telescope (LAT) on the \emph{Fermi} Gamma-ray Space Telescope (formerly GLAST). This discovery makes PSR J0030+0451 the second millisecond pulsar to be detected in gamma-rays after PSR J0218+4232, observed by the EGRET instrument on the Compton Gamma Ray Observatory. The spin-down power $\dot E = $ 3.5 $\times$ 10$^{33}$ ergs s$^{-1}$ is an order of magnitude lower than the empirical lower bound of previously known gamma-ray pulsars. The emission profile is characterized by two narrow peaks, respectively 0.07 $\pm$ 0.01 and 0.08 $\pm$ 0.02 wide, separated by 0.44 $\pm$ 0.02 in phase. The first gamma-ray peak falls 0.15 $\pm$ 0.01 after the main radio peak. The pulse shape is similar to that of the ``normal'' gamma-ray pulsars. An exponentially cut-off power-law fit of the emission spectrum leads to an integral photon flux above  100 MeV of (6.76 $\pm$ 1.05 $\pm$ 1.35) $\times 10^{-8}$ cm$^{-2}$ s$^{-1}$ with cut-off energy (1.7 $\pm$ 0.4 $\pm$ 0.5) GeV. Based on its parallax distance of $(300 \pm 90)$ pc, we obtain a gamma-ray efficiency $L_\gamma / \dot{E} \simeq 15\%$ for the conversion of spin-down energy rate into gamma-ray radiation, assuming isotropic emission.
\end{abstract}
\keywords{pulsars:general -- pulsars:individual (PSR J0030+0451) -- gamma-rays:observations}

\section{Introduction}

Two distinct pulsar populations are the ``normal'' and ``millisecond'' pulsars (MSPs),  the latter being rapidly rotating neutron stars ($P \lesssim$ 30 ms) with very small period increases ($\dot{P} \lesssim$ 10$^{-17}$ s/s). MSPs represent roughly 10\% of the pulsars listed in the ATNF online catalogue \citep{manchester2005}.  In the classical framework of the magnetic braking model, MSPs are old stars with characteristic spin-down ages $\tau = P/(2 \dot{P}) \simeq $ (0.1 -- 10) $\times$ 10$^{9}$ yrs, and characteristic surface dipole magnetic fields, B$_{surf} \simeq $ 3.2 $\times$ 10$^{19}$ $(P \dot{P})^{1/2} <$ 10$^{10}$ G. Most MSPs are in binary systems. They are thought to have been spun-up by the accretion of matter and thus transfer of angular momentum from a binary companion \citep{alpar92}. In some binary systems, the companion is evaporated by the strong relativistic wind produced by the millisecond pulsar \citep{ruderman89}, which becomes isolated, like PSR J0030+0451. 

PSR J0030+0451 was discovered by two independent radio surveys, the Arecibo Drift Scan Search \citep{somer2000} and the Bologna sub millisecond pulsar survey \citep{damico2000}. Its spin-down age $\tau$ is 7.6 $\times$ 10$^{9}$ yrs. The analysis of radio timing residuals showed a significant annual parallax of 3.3 $\pm$ 0.9 mas, leading to a distance measurement of 300 $\pm$ 90 pc  \citep{lommen2006}, useful for luminosity estimates. The \citet{cordeslazio2002} model of Galactic electron distribution predicts a distance of 317 pc, in good agreement with the parallax measurement. In addition, \citet{lommen2006} also argued that the Shklovskii effect \citep{shklov70} on the pulsar's first period derivative $\dot P$ is less than 1\%, hence rejecting the possible contamination of proper motion in $\dot P$ and reinforcing the determination of the spin-down energy rate $\dot E$. This $P = 4.87$ ms  and $\dot P = 1.0 \times 10^{-20}$ s s$^{-1}$ pulsar hence has an $\dot E = 4 \pi^{2} I (\dot P / P^3)$ of 3.5 $\times$ 10$^{33}$ ergs s$^{-1}$,  taking the moment of inertia $I$ to be 10$^{45}$ g cm$^2$. The magnetic field strength at the stellar surface $B_S = 3.2 \times 10^{19}$ G $\sqrt{P \dot P}$ for this pulsar is 2.2 $\times$ 10$^8$ G, and its characteristic age $\tau = P / (2 \dot P)$ is 7.7 $\times$ 10$^9$ yrs.

This pulsar was detected in X-rays with \emph{ROSAT}, during the final days of the mission, shortly after its discovery in radio
\citep{becker2000}. The MSP was then observed by XMM-\emph{Newton} \citep{beckasch2002}. Both telescopes revealed a broad X-ray pulsation profile,  with a pulsed fraction compatible with 50\% and the two X-ray peaks separated by 0.5 in phase. Unfortunately neither of the two X-ray telescopes could provide the X-ray alignment relative to the radio at the time of the observations, due to the lack of accurate clock calibration. We address this alignment issue below. \citet{beckasch2002} showed that the X-ray spectrum is consistent with being purely thermal. \citet{bogdanov2008} invoked the presence of a hydrogen atmosphere to model the X-ray spectrum.

Six pulsars were detected in gamma-rays with high confidence with the EGRET telescope \citep{djt99}, and more recently AGILE and LAT reported the detection of pulsed gamma-rays from PSR J2021+3651 \citep{halpern2008, Fermi2021}. The LAT also discovered a young pulsar in the supernova remnant CTA1 \citep{FermiCTA1}. All eight are normal pulsars. Gamma-ray pulsations with 4.9 $\sigma$ statistical significance were reported for the 2.3 ms pulsar J0218+4232 \citep{kuip2004} with EGRET. No pulsed gamma-ray emission had previously been detected from an MSP \citep{fierro95}. LAT observations confirm the detection of gamma-ray emission from PSR J0218+4232 \citep{FermiMSPs}. 

The revised EGRET catalogue of gamma-ray sources of \citet{casgren2008} lists the new source EGR J0028+0457 with no 3EG counterpart, with a 95\% confidence contour radius of 0.51$\degr$. The millisecond pulsar PSR J0030+0451, for which \citet{harding2005} predicted a gamma-ray flux superior to that of the marginal EGRET detection PSR J0218+4232, is located 0.5$\degr$ from EGR J0028+0457 and was hence suggested as a possible counterpart. The launch of the \emph{Fermi} observatory, formerly GLAST, on 2008 June 11 provided a new opportunity to study the emission of gamma-rays by MSPs. 

This paper describes the discovery of pulsations from the isolated pulsar PSR J0030+0451 in the \emph{Fermi}  Large Area Telescope (LAT) data, making this the first high-confidence detection of a millisecond pulsar in gamma-rays. We also discuss the timing analysis of  XMM-\emph{Newton} data with accurate clock calibration, providing the phase alignment of the radio, X-ray and gamma-ray emission relative to each other.  

\section{Observations and analysis}

The LAT instrument is described in \citet{glast}. Gamma-rays convert to electron-positron  pairs in the tracker, consisting of tungsten foils interleaved with layers of silicon microstrip detectors. The tracker provides direction information. Below the tracker is the calorimeter in which particles interact in cesium iodide crystals, giving most of the energy information. The detector is surrounded by the anticoincidence detector, which helps reject the charged cosmic-ray background. The LAT is sensitive to photons with energies  below 20 MeV to over 300 GeV. Its large field of view of 2.4 sr, large effective area of 8000 cm$^2$ on-axis at 1 GeV, improved angular resolution (0.5$\degr$ of  68\% PSF containment at 1 GeV for events collected in the `front' section with thin radiator foils), the scanning observing mode, and small trigger deadtime of 26.5 $\mu$s, make the LAT much more sensitive than EGRET. Ground tests using cosmic ray muons demonstrated that the LAT measures event times with a precision better than 1 $\mu$s. On orbit satellite telemetry indicates comparable accuracy. The software timing chain from the GPS-based satellite clocks through the barycentering and phase-folding software has been shown to be accurate to better than a few $\mu$s \citep{smith2008}. 

\begin{figure}
\includegraphics[angle=270,scale=0.33]{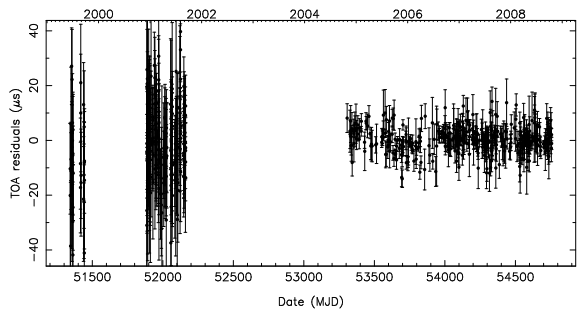}
\caption{Timing residuals as a function of time for the model given in Table \ref{tab1}. Data between 1999 and 2002 were recorded using the NBPP backend, the BON backend was used after 2004.\label{residus}}
\end{figure}

PSR J0030+0451 is a very stable pulsar. The timing solution used for this object has been derived from observations made with the
Nan\c{c}ay radiotelescope, near Orleans, France. About seven hundred observations starting in July 1999 were used, contemporaneous with the \emph{Fermi} LAT dataset and bracketing the XMM observation. Between 1999 and 2002, data were recorded with the Navy Berkeley Pulsar Processor (NBPP) backend. This instrumentation was designed and built at the Naval Research Laboratory in collaboration with the University of California, Berkeley \citep{backer97,foster96}. The system covers 1.5 MHz per channel for a total bandwidth of 144 MHz centered at 1360 MHz. In timing mode, the data were folded for 15 minutes over 256 bins covering the full period of the pulsar PSR J0030+0451 (with a resolution of $\simeq$ 20 $\mu$s). Since late 2004, TOAs are obtained from the coherent pulsar instrumentation currently in use at Nan\c{c}ay, the Berkeley-Orl\'{e}ans-Nan\c{c}ay (BON) backend \citep{theureau2005,cognard2006}. The coherent dedispersion is performed within 4 MHz channels over a total bandwidth of 64 MHz (128 MHz since July 2008) centered at 1398 MHz. Compared to those recent high quality data, the old NBPP TOAs were degraded due to the absence of dedispersion inside the 1.5 MHz channels (also producing large systematics when the pulsar scintillates). The bulk of radio observations were done at 1.4 GHz (1360 $\pm$ 72 MHz before 2002, 1398 $\pm$ 32 MHz after 2004 and 1398 $\pm$ 64 MHz after 2008). 2 GHz observations were made to better constrain the dispersion measure, necessary for the alignment of pulsar profiles at different wavelengths. The TEMPO\footnote{\url{http://www.atnf.csiro.au.research/pulsar/tempo}} package was used to build a timing solution from the recorded radio times of arrival, determined through a standard cross-correlation procedure \citep{taylor92}. The mean time of arrival uncertainty is 3.6 $\mu$s for the recent BON observations (the bulk of the data), while this is 8.6 $\mu$s for all the dataset. The post-fit rms is 3.7 $\mu$s, fitting for the pulsar's frequency and first-derivative, and taking its proper motion into account. The timing parameters used for phase-folding gamma-ray events are given in Table \ref{tab1}, the timing residuals after the fitting procedure are plotted in Figure \ref{residus}. Doubling the error in the parallax given by TEMPO, and adding in quadrature an additional 0.3 mas uncertainty due to the solar wind, as explained in \citet{lommen2006}, we measure a parallax of 4.1 $\pm$ 0.7 mas, which is consistent with the Lommen et al value. The dispersion measure we derive is also consistent with the value of 4.3328(3) pc cm$^{-3}$ quoted in \citet{lommen2006}. 

\begin{deluxetable}{ll}
\tablewidth{0pt}
\tabletypesize{\scriptsize}
\tablecaption{Timing ephemeris for pulsar PSR J0030+0451.\label{tab1}}
\tablehead{\colhead{Parameter} & \colhead{Value}}
\startdata
Right ascension, $\alpha$\dotfill & 00:30:27.4303(5) \\
Declination, $\delta$\dotfill & 04:51:39.74(2) \\
Proper motion in right ascension, $\mu_\alpha$ (mas yr$^{-1}$)\dotfill & -5.3(9) \\
Proper motion in declination, $\mu_\delta$ (mas yr$^{-1}$)\dotfill & -2(2) \\
Parallax (mas)\dotfill & 4.1(3) \\
Epoch of position determination (MJD)\dotfill & 52079 \\
Pulse frequency, $\nu$ (s$^{-1}$)\dotfill & 205.530699274922(9) \\
First derivative of pulse frequency, $\dot{\nu}$ (s$^{-2}$)\dotfill & -4.2976(4) $\times$ 10$^{-16}$ \\
Epoch of ephemeris (MJD)\dotfill & 50984.4 \\
MJD range\dotfill & 51343 -- 54757 \\
Number of TOAs\dotfill & 651 \\
Rms timing residual ($\mu s$)\dotfill & 3.69 \\
Dispersion measure, $DM$ (cm$^{-3}$pc)\dotfill & 4.333(1) \\
Solar system ephemeris model\dotfill & DE200 \\
\enddata
\tablecomments{Figures in parentheses are the nominal 1$\sigma$ \textsc{tempo} uncertainties in the least-significant digits quoted. 
Epochs are given in TDB units.}
\end{deluxetable}

The LAT data considered here were taken during \emph{Fermi}'s first-year all-sky survey, starting 2008 August 3, through November 2. The pulsar is located  well outside of the Galactic plane (l = 113.141$\degr$ and b = -57.611$\degr$) and is hence in a region of low Galactic background. Aiming to have a good signal-to-noise ratio over a broad energy range, we used an energy-dependent region of interest of $\theta =$ Max$[0.9-2.1\ \rm{Log}_{10}(E_{GeV}), 1.5]$ degrees around the pulsar position. A larger fraction of the PSF is included at high energies, where there is little background contamination. Since \emph{Fermi} was operating in survey mode,  the contribution of the Earth's gamma-ray albedo to the background was  negligible. Finally, the ``diffuse'' class events were kept. A description of event classes can be found in \citet{glast}. After application of the above cuts, we obtained a dataset of 563 events over 100 MeV.

PSR J0030+0451 was observed by XMM-\emph{Newton} on 2001 June 19-20. The observations with the pn camera spanned 29 ks, but a soft proton flare affected approximately 8.8 ks of the  exposure.  The data were reduced using Version 8.0 of the {\it XMM-Newton} Science Analysis Software\footnote{\url{http://xmm.vilspa.esa.es/sas/}}. This version solves the timing problem. The pn camera was used in timing mode with the thin filter. The data were reduced using ``epproc''.  The event lists were filtered, so that 0 -- 4 of the predefined patterns (single and double events) were retained, as these have the best energy calibration. We used the data between 0.3 and 2.5 keV as this had the best signal-to-noise. The event times were converted to Barycentric Dynamical Time, using the task ``barycen'' and the same coordinates as used for the {\em Fermi} data. The XMM-\emph{Newton} absolute timing accuracy is found to be as good as 300 $\mu$s\footnote{\url{http://xmm2.esac.esa.int/docs/documents/CAL-TN-0045-1-0.pdf}}, or $\simeq$ 0.06 rotations of PSR J0030+0451. In order to obtain such fine temporal resolution with the pn in timing mode, the CCD is read out continously, causing the events for the target source to be smeared out in the Y-direction. To extract all the events from the source, we used the standard procedure of creating a one dimensional image, by binning all of the raw data in the Y-direction into a single bin. The spectrum was extracted using a range of 7 pixels centered on the pulsar, in the X-direction. The background spectrum was extracted from a similar neighbouring region, free from X-ray sources.



\section{Results}

\subsection{Gamma-ray and X-ray profile analysis}

\begin{figure}
\epsscale{1.}
\plotone{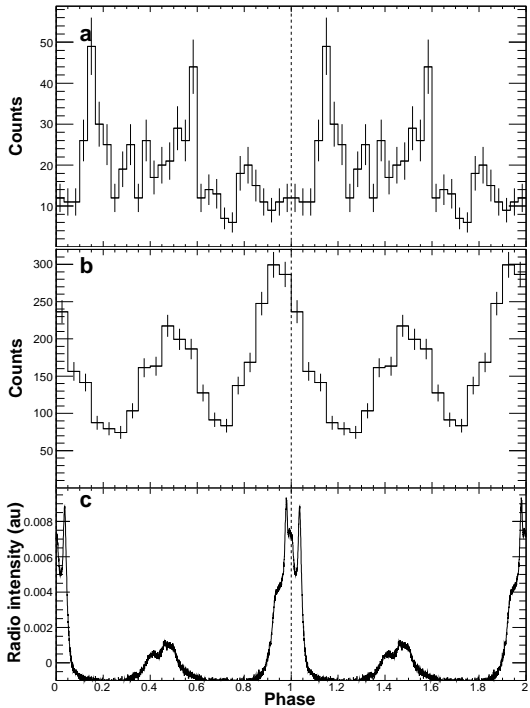}
\caption{Multi-wavelength phase histograms of PSR J0030+0451. Two pulsar rotations are shown. a: Gamma-ray phase histogram of PSR J0030+0451 at E $>$ 100 MeV, within an energy-dependent ROI. Each bin is 0.033 in phase, or 160 $\mu$s. b: 20 bin 0.3--2.5 keV XMM-\emph{Newton} phase histogram. c: Radio profile obtained at 1.4 GHz used to build the ephemeris. \label{phaso}}
\end{figure}

Figure \ref{phaso} shows the phase histogram of the events with energies greater than 100 MeV in panel a, along with the reference radio profile used to derive the timing ephemeris in panel c. The $\chi^2$ value for the phase histogram shown in panel a of Figure \ref{phaso} is 121 for 29 degrees of freedom, indicating that the probability that the pulsation is actually a statistical fluctuation is 3.1 $\times$ 10$^{-13}$. The bin independent \emph{H}-Test \citep{dejager89} gives a value of 123. The derived chance occurence probability of the null hypothesis, non-pulsed emission, is
below 4 $\times$ 10$^{-8}$. The null hypothesis is hence ruled out, for a single trial. 

The zero of phase in the reference radio profile is defined to be at the maximum of the first Fourier harmonic of the signal, transferred back to the time domain. The maximum of the main radio peak's second subpulse is at 0.036 in phase. The gamma-ray pulse profile comprises two peaks (see Figure \ref{phaso}, panel a). There is a shift between the first gamma-ray peak (P1), occuring at 0.15 $\pm$ 0.01 in phase, and the main radio peak. The gamma-ray peaks of PSR J0030+0451 are very sharp. We fit P1 and P2 with two-sided lorentzians, to take into account the different widths for the leading and trailing edges. For the first peak, the fit gives a full width at half maximum of 0.07 $\pm$ 0.01 in phase, that is, 340 $\mu$s. The second gamma-ray peak, P2, lags P1 by $\Delta\Phi = $ 0.44 $\pm$ 0.02 in phase. The fit of the structure between 0.45 and 0.6 places the peak at 0.59 $\pm$ 0.01 in phase with a full width at half maximum (FWHM) of 0.08 $\pm$ 0.02 in phase. 

\begin{figure}
\epsscale{1.}
\plotone{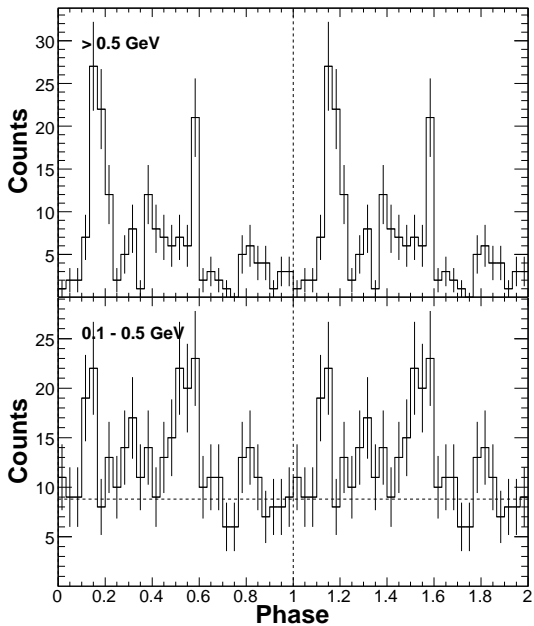}
\caption{30 bin phase histograms for PSR J0030+0451 with the LAT, in different energy bands. Two rotations are shown. The horizontal dashed line shows the background level estimated from a surrounding annulus.\label{phaso2}}
\end{figure}

LAT phase histograms in two energy bands are given in Figure \ref{phaso2}, with 30 bins per rotation. The pulsar is faint in the 100 to 500
MeV band. In this energy band, P2 is prominent relative to P1: taking P1 (resp. P2) between 0.1 and 0.25 in phase (resp. 0.45 and 0.60), the P1/P2 ratio is found to be 0.47 $\pm$ 0.17. The upper panel of Figure \ref{phaso2} shows the events with energies over 500 MeV. In this band P1 dominates P2, with a ratio of 1.64 $\pm$ 0.48. There hence seems to be a spectral dependence of the gamma-ray profile. Within error bars, the P1/P2 ratio seems to increase as a function of energy, conversely to the Vela pulsar \citep{velapap}. The first gamma-ray peak hence seems harder than the second one. More photons are needed to perform phase-resolved spectroscopy. We note that in the 100 MeV to 500 MeV energy band, P2 seems broader and closer to P1 than it does for energies greater than 500 MeV. However, we cannot be conclusive about this trend due to the low number of photons in P2: more statistics might reveal bridge emission between P1 and P2, in which case P2 would be narrower than the 0.08 +/- 0.02 value quoted above. We also note that there is no evidence for pulsed emission from 20 MeV to 100 MeV.  Extrapolation of the hard, observed spectrum (see spectral analysis below) predicts only a few photons in this energy band, and any pulsation is almost certainly obscured by the rapidly rising background below 200 MeV.


The overall gamma-ray emission profile is reminiscent of the younger pulsars Vela, Crab, Geminga, PSR B1951+32 or PSR J2021+3651, especially over 500 MeV. The phase separation of $\simeq$ 0.4 between the two peaks is a common feature \citep{djt2004}. The pulse profile also shows a main gamma-ray peak lagging the main radio component by 0.15 $\pm$ 0.01, similar to the main gamma peak lagging 0.11 to 0.16 after the radio pulse for Vela, PSR B1951+32 and PSR J2021+3651. The radio to P1 and P1 to P2 separation are in agreement with the outer-gap pulse profile model of \citet{romyad95}, which compares the gamma-ray peak separation $\Delta$ with the radio to gamma-ray lag $\delta$ for a given magnetic inclination angle $\alpha$. This is also in agreement with the two-pole caustic and slot gap models for gamma-ray emission \citep{watters2008, dykshard2004}. With radio polarization measurements, \citet{lommen2000} found the magnetic inclination angle $\alpha$ to be most probably 62$\degr$, that is a mostly  orthogonal configuration.

Panel b in Figure \ref{phaso} shows the 0.3 to 2.5 keV XMM-\emph{Newton} phase histogram for PSR J0030+0451. The overall profile is similar to
that found by \citet{becker2000} and \citet{beckasch2002}, though their observations did not yield the X-ray  to radio alignment because of inaccurate absolute timing. Here the $\simeq$ 300 $\mu$s absolute timing accuracy represents $\pm$ 1 bin in panel b. The X-ray and radio are hence consistent with being phase aligned. This result supports the idea that the X-ray and radio emission have common origins in the pulsar magnetosphere, and that gamma-rays are produced in a different region. In that sense, PSR J0030+0451 is different from PSR J0218+4232: according to \citet{kuip2000}, the 2.3 ms pulsar has its radio, X-ray and gamma-ray components aligned.

\subsection{Gamma-ray spectral analysis}

Using the standard maximum-likelihood spectral estimator \emph{gtlike} in the \emph{Fermi} Science Tools\footnote{\url{http://fermi.gsfc.nasa.gov/ssc/data/analysis/SAE\_overview.html}}, we performed a spectral analysis of the gamma sample. The diffuse Galactic background and extragalactic emissions are taken into account, as well as the instrument response, which is a function of the photon energy and incidence angle relative to the telescope axis. The dataset used for the spectral analysis spans the same time interval as for the timing analysis, but this time we retain events within 15\degr \ of the millisecond pulsar. Because of present uncertainty in the instrument response below 200 MeV, events with energy below 200 MeV are rejected.

\begin{figure}
\plotone{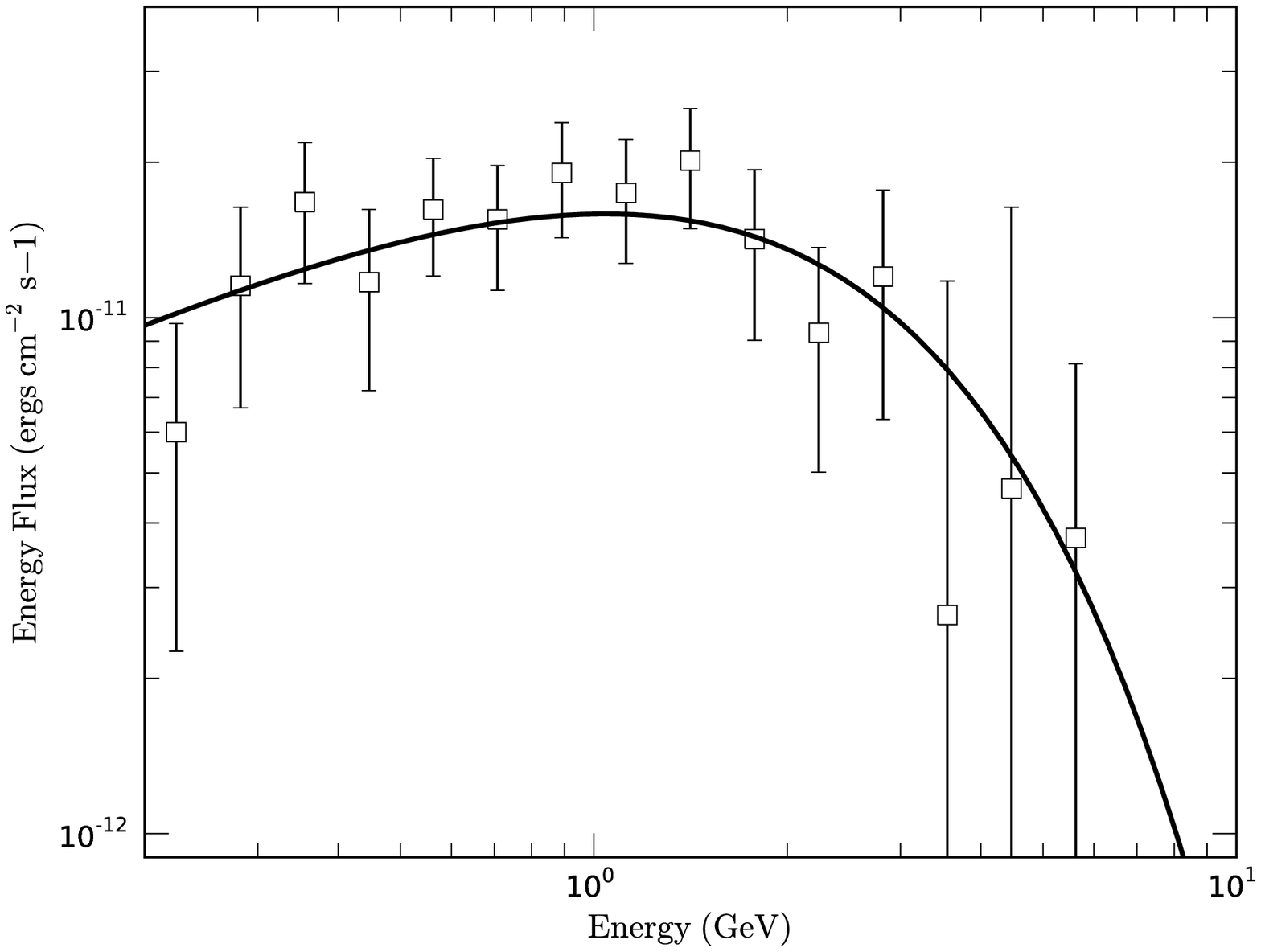}
\caption{Phase-averaged spectral energy distribution for PSR J0030+0451, along with the best-fit power law with exponential cut-off. Statistical errors are shown.\label{spectrum}}
\end{figure}

Figure \ref{spectrum} shows the phase-averaged differential energy spectrum. The corresponding power-law with exponential cutoff modeling the data is given by:

\begin{eqnarray}
\frac{dN_\gamma}{dE} = N_0 \left(\frac{E}{10^3\ \rm{MeV}}\right)^{- \Gamma} e^{- E / E_c}
\label{eqspectre}
\end{eqnarray}

In this expression, $E$ is in MeV, the prefactor term $N_0 = $ (1.84 $\pm$ 0.38 $\pm$ 0.37) $\times 10^{-11}$ $\rm{ph}\ \rm{cm}^{-2}\ \rm{s}^{-1}\ \rm{MeV}^{-1}$, the power-law index $\Gamma = $ (1.4 $\pm$ 0.2 $\pm$ 0.2) and the cut-off energy $E_c = $ (1.7 $\pm$ 0.4 $\pm$ 0.5) GeV. The first error is statistical, the second is systematic, dominated by differences between the a priori expectations of the \emph{Fermi} LAT effective area and the on-orbit instrument response. The magnitude of this effective area uncertainty is $<$ 10\% near 1 GeV, 20\% below 0.1 GeV and 30\% over 10 GeV. Analysis improvements are underway.

The stability of the results was tested by fitting the same dataset with a binned maximum likelihood estimator, `ptlike', which computes the photon counts in a point source weighted aperture in excess of background counts. The fit results, shown in Figure \ref{spectrum} for each energy band, are consistent with those obtained with `gtlike' within error bars. We have also tried a simple power-law fit to the data, of the form $dN_\gamma / dE = N_0 (E/ 1 \rm{GeV})^{- \Gamma}$. The fit using the exponential cutoff functional form is better constrained, with a difference in the log likelihoods of 10.76. A $\chi^2$ interpretation of this value leads to a probability of incorrectly rejecting the power-law hypothesis of 3.5 $\times$ 10$^{-6}$.

Integrating Equation (\ref{eqspectre}) for energies $>$ 100 MeV yields an integral flux $f_{> 100 MeV} = $ (6.76 $\pm$ 1.05 $\pm$ 1.35) $\times 10^{-8}$ $\rm{ph}\ \rm{cm}^{-2}\ \rm{s}^{-1}$. The revised EGRET catalogue of \citet{casgren2008} quotes a flux of (10.4 $\pm$ 3.1) $\times$ 10$^{-8}\ \rm{ph}\ \rm{cm}^{-2}\ \rm{s}^{-1}$ for EGR J0028+0457, based on the summed EGRET dataset. Both fluxes are statistically in agreement. The pulsar is in a low EGRET exposure region, mainly detected at large off-axis angles where systematic uncertainties in the effective area were large.

The energy flux is $F_{obs} = $ (4.91 $\pm$ 0.45 $\pm$ 0.98) $\times 10^{-11}$ $\rm{ergs}\ \rm{cm}^{-2}\ \rm{s}^{-1}$ over 100 MeV. The luminosity of a pulsar can be written as $L_\gamma = 4 \pi f_\Omega F_{obs} D^2$, where $D$ is the pulsar distance, $f_\Omega$ is a correction factor containing information about the beaming geometry, and $F_{obs}$ is the observed phase-averaged energy flux. \citet{watters2008} have computed pulse profiles and $f_\Omega$ corrections for young pulsars. The outer magnetospheres for MSPs should be scaled-down analogues of those of young pulsars. If we use the maps in \citet{watters2008} for a pulse separation $\Delta = 0.4$ and a magnetic inclination $\alpha=62^\circ$, as inferred from the radio polarization data, we find a reasonable match for an old, high efficiency two-pole caustic/slot gap model for a viewing angle $\zeta \simeq 75^\circ$. In turn this implies $f_\Omega \simeq 0.8$. This pulse width is not natural in a high efficiency, large gap width outer gap model, suggesting that the lower altitude two-pole caustic picture is a better match to the data, unless the true pulsar efficiency is $\la 0.1$. Adopting $f_\Omega = 1$ for PSR J0030+0451, we obtain an efficiency $\eta = L_\gamma / \dot{E} = $ 15\% for the conversion of spin-down energy into gamma-ray emission.

\citet{arons96} noted that for the EGRET pulsars, $\eta$ is, in good approximation, inversely proportional to the open field line voltage $V = 4 \times 10^{20} P^{-3/2} \dot{P}^{1/2}$, proportional to $\sqrt{\dot{E}}$, and also proportional to the open field current \citep{harding81}. With an open field line voltage of 1.2 $\times$ 10$^{14}$ volts and an efficiency of 15\%, PSR J0030+0451 seems to break from the trend, which would predict a higher value. However, the efficiency law may saturate at lower spin-down rates. Other low $\dot E$ pulsar detections with \emph{Fermi} should help constrain how these old ``recycled'' pulsars convert their energy loss rate into gamma-ray luminosity. In addition, the discovery of pulsed emission from PSR J0030+0451, which has a smaller spin down rate than previously known gamma-ray pulsars, lowers the empirical $\dot E$ threshold for gamma-ray emission by an order of magnitude. This suggests that many low $\dot E$ pulsars might be detectable by the LAT.

Attempts to describe the high-energy emission from MSPs were recently made based on the two main classes of theoretical models: the polar cap (PC) and the outer gap (OG) models. In the PC description by \citet{harding2005}, charged particles are accelerated along the open field lines near the magnetic poles to high altitudes. The high-energy spectrum consists of three main components. The emission of photons up to 100 MeV is dominated by synchrotron radiation from electrons.  Over 100 GeV, photons are produced by inverse Compton radiation from electrons. Curvature radiation from electrons dominates the photon spectrum between 1 and 100 GeV. For PSR J0030+0451, \citet{harding2005} provide a prediction of 4.25 GeV for the curvature radiation cutoff energy. This value differs from the 1.7 GeV we find in this analysis. In their model, the expected curvature radiation flux over 100 MeV is $F_{CR}(>100 \rm{MeV}) \sim 3 \times 10^{-6} \rm{ph}\ \rm{cm}^{-2}\ \rm{s}^{-1}$ for our pulsar. Our measured integral flux disagrees with their expectation. However, \citet{harding2005} computed sky-averaged spectra; more detailed three-dimensional sums may be needed to make predictions for an individual viewing angle. On the other hand, in the OG description of the high-energy radiation from MSPs by \citet{zhang2003}, a strong multipole magnetic field exists near the stellar surface. X-rays are then produced by the backflow current of the outer gap, which is a vacuum gap close to the light cylinder. These X-rays consist of a non-thermal power-law component, plus two thermal components. Gamma-rays are produced in the outer gap. Though \citet{zhang2003} provide no application of their model to the case of PSR J0030+0451, they predict that a gamma-ray MSP would be an X-ray MSP. This is indeed the case for both PSR J0030+0451 and PSR J0218+4232. Secondly, they predict that if the X-ray spectrum is dominated by thermal emission, which is indeed the case for PSR J0030+0451, consistent with being purely thermal \citep{beckasch2002}, the gamma-ray emission can extend to $\sim$ GeV gamma-rays only, as we observe.

\section{Conclusion}

We described the detection of the millisecond pulsar PSR J0030+0451 in gamma-rays using the \emph{Fermi} LAT. We now have high confidence that there are gamma-ray emitters among millisecond pulsars. This provides a new tool for studying the magnetospheres of energetic pulsars. Detection of more millisecond pulsars in the LAT data may invite revisiting the possible contribution of unresolved millisecond pulsars to the overall Galactic diffuse emission in the gamma-ray band: for instance \citet{wang2005} proposed that the Galactic centre might contain a few thousand unresolved MSPs, contributing to the diffuse spectrum detected by EGRET, which shows a break at a few GeV. One might also expect cumulative gamma-ray emission of millisecond pulsars in globular clusters, such as 47 Tuc, which is thought to contain up to 60 MSPs \citep{camilo2005}. The EGRET search for emission from globular clusters only provided upper limits \citep{michelson94,fierro95}. The \emph{Fermi} LAT offers new opportunities to search for emission from globular clusters in gamma-rays.

\acknowledgments

The  \emph{Fermi}  LAT Collaboration acknowledges generous ongoing support from a number of agencies and institutes that have supported both the development and the operation of the LAT as well as scientific data analysis.  These include the National Aeronautics and Space Administration and the Department of Energy in the United States, the Commissariat \`{a} l'Energie Atomique and the Centre National de la Recherche Scientifique / Institut National de Physique Nucl\'{e}aire et de Physique des Particules in France, the Agenzia Spaziale Italiana and the Istituto Nazionale di Fisica Nucleare in Italy, the Ministry of Education, Culture, Sports, Science and Technology (MEXT), High Energy Accelerator Research Organization (KEK) and Japan Aerospace Exploration Agency (JAXA) in Japan, and the K.~A. Wallenberg Foundation, the Swedish Research Council and the Swedish National Space Board in Sweden.  

Additional support for science analysis during the operations phase from the following agencies is also gratefully acknowledged: the Istituto Nazionale di Astrofisica in Italy and the K.~A. Wallenberg Foundation in Sweden for providing a grant in support of a Royal Swedish Academy of Sciences Research fellowship for JC.

\end{document}